\documentclass[sigconf,9pt, authorversion]{acmart}

\copyrightyear{2020} 
\acmYear{2020} 
\setcopyright{acmcopyright}
\acmConference[SOSR '20]{Symposium on SDN Research}{March 3, 2020}{San Jose, CA, USA}
\acmBooktitle{Symposium on SDN Research (SOSR '20), March 3, 2020, San Jose, CA, USA}
\acmPrice{15.00}
\acmDOI{10.1145/3373360.3380831}
\acmISBN{978-1-4503-7101-8/20/03}

\usepackage{multirow}
\usepackage{amsmath}
\usepackage{xcolor}
\usepackage{mathtools}
\usepackage{xspace}
\usepackage{tikz}
\usepackage{tikz-qtree}
\usepackage{amssymb}
\usepackage{booktabs}
\usepackage{caption}
\usepackage{subcaption}
\usepackage{listings}

\newcommand{\name}{Anime\xspace}
\newcommand{\hre}{HRE\xspace}
\newcommand{\hres}{HREs\xspace}

\newcommand{\intent}{intent\xspace}
\newcommand{\intents}{intents\xspace}

\newcommand{\Intents}{Intents\xspace}

\begin{document}

\title{Automatic Inference of High-Level Network \Intents \\by Mining Forwarding Patterns}

\renewcommand{\shorttitle}{Automatic Inference of High-Level Network \Intents by Mining Forwarding Patterns}

\author{
Ali Kheradmand
}
\affiliation{
    \institution{
    University of Illinois at Urbana-Champaign
    }
}


\begin{abstract}
There is a semantic gap between the high-level intents of network operators and the low-level configurations that achieve the intents. 
Previous works tried to bridge the gap using verification or synthesis techniques, both requiring formal specifications of the intended behavior which are rarely available or even known in the real world. 
This paper discusses an alternative approach for bridging the gap, namely to infer the high-level intents from the low-level network behavior. Specifically, we provide Anime, a framework and a tool that given a set of observed forwarding behavior, automatically infers a set of possible intents that best describe all observations. Our results show that Anime can infer high-quality intents from the low-level forwarding behavior with acceptable performance. 
\end{abstract}





\maketitle

\section{Introduction}
\label{sec:intro}
As a computer network becomes more complex over time, its correctness becomes an increasingly important concern for the organization operating it. 
The way most of today's networks are configured is that given an informal high-level description of how the network should operate, a network administrator configures the network devices to achieve that. These descriptions implicitly define networking \textit{\intents}. \Intents are usually network-wide properties of the network forwarding behavior and are simple enough to be expressed and comprehended by humans, e.g. the traffic received from the Internet destined to IP $x$ must reach node $A$, or the SSH traffic from node $B$ to node $C$ must go through a DPI device and be resilient to any 2-link failures. 
The administrator converts these high-level \intents into low-level device configurations, often manually. 
Consequently, there is a significant gap between the high-level \intents of network operators and the low-level configurations~\cite{propane}. 
This gap is a source of many misconfigurations and network problems, leading to catastrophic consequences including network outages and breaches that often make news headlines~\cite{propane}.


As a result, there has been significant research progress towards network verification tools~\cite{veriflow, deltanet, pec, p4k, plankton, minesweeper, plankton20} that given a set of network-wide \intents, check whether the configured network satisfies the \intents. 
There has also been progress towards network programming/configuration synthesis~\cite{netkat, frenetic, propane, pga} tools that given the \intents described in a domain-specific language, synthesize data plane entries or control plane configurations that satisfy the \intents. 

These tools rely on the ability of an administrator to provide a formal specification of the desired behavior. 
However, such specifications do not usually exist in practice for current networks.  
Administrators often inherit an already working legacy network and are asked to maintain it. 
In the real world, administrators hesitate to touch the network they operate very often due to concerns over breaking the network~\cite{ddnf}.
Such networks being so complex, it is practically challenging to manually identify the unwritten high-level \intents by looking at the low-level configurations.

In this paper, we take an alternative approach to bridge the gap.
We present \name (\textit{Automatic Network Intent Miner}), a framework and a prototype tool to infer high-level \intents by mining the commonalities among the forwarding behavior in the network.
The output can be used for human comprehension, continuous verification and anomaly detection, automatic migration to SDN, etc.



After introducing objective measures of quality for intent inference, we accordingly formalize the intent inference problem in a framework that fits the hierarchical nature of networks such as in topological and address hierarchies. We then provide a heuristic solution to the problem based on clustering techniques.
The result, \name, takes as input a set of forwarding behavior observed in one or more snapshots of the network expressed using various features such as the packet header information, devices along the path, time of path observation, device or link state, etc.
Given a limit on the number of inferred intents, \name produces a set of intents that collectively describe all observed behavior with high precision. The results also predict unobserved but possible behavior.





We evaluate the effectiveness and performance of \name for a use case where the goal is to only summarize observed network behavior for human comprehension. We also evaluate the tool in settings where not all possible behavior is observed and some needs to be predicted. 
As a baseline, we compare the results to the closest related work (Net2Text\cite{net2text}), though the goal of that work is not intent inference (Section~\ref{sec:related-work}).
The results demonstrate \name's ability in inferring higher quality \intents with acceptable performance.

\newpage
\section{Motivation}
In this section, we show how \intents can be inferred from low-level forwarding behavior and discuss the applications of doing so.

\subsection{Illustrative Examples}
\label{sec:example}
We provide two examples illustrating how forwarding behavior can be used to infer possible \intents. The examples are inspired by the ones in network verification literature~\cite{propane, netgen, deltanet}. In these examples, the information from forwarding paths across various devices, packet headers, and data plane snapshots are used to derive higher-level information about the collection of the paths.  

\subsubsection{Example 1: Data center network}

\newcommand{\IPa}{10.0.1.2}
\newcommand{\IPb}{10.0.1.3}

\begin{figure}[t!]
    \centering
    \captionsetup[subfigure]{aboveskip=-1pt,belowskip=-1em}
    \begin{subfigure}[t]{1in}
        \begin{center}
        \includegraphics{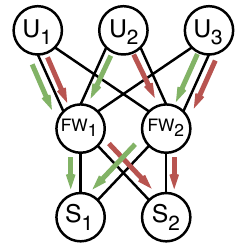}
        \caption{Example 1}
        \label{fig:dc}
        \end{center}
    \end{subfigure}%
    ~
    \begin{subfigure}[t]{2.2in}
        \begin{center}
        \includegraphics[width=0.85\textwidth]{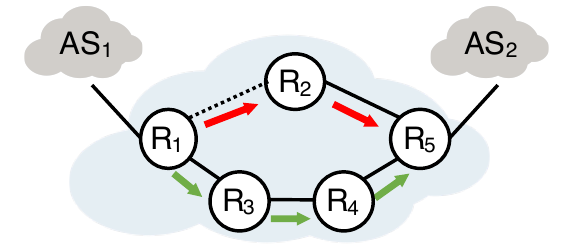}
        \caption{Example 2}
        \label{fig:as}
        \end{center}
    \end{subfigure}
    \caption{Example network setups}
    \vspace{-1em}
\end{figure}


Consider the network in Figure~\ref{fig:dc} resembling part of a very simple data center network.
The network contains three user machines ($U_1$, $U_2$, $U_2$), two firewalls ($FW_1$, $FW_2$), and two servers ($S_1$, $S_2$). 
The green and red arrows in the figure denote the forwarding paths for packets with destination IP addresses $\IPa$ and $\IPb$ respectively.

Let's consider two out of the three green paths destined to $\IPa$, $U_1 . FW_1 . S_1$ and $U_2 . FW_1 . S_1$. The only difference between the two paths is their starting node ($U_1$ vs. $U_2$). Looking at these paths we could say that the \intent of the administrator is to enforce all packets starting from the user nodes destined to $\IPa$ to go through firewall $FW_1$ and reach server $S_1$. 
We represent this guess as: $dstIP:\IPa, start:User, waypoint:FW_1, end:S_1$.

Let us now consider the third green path, namely $U_3.FW_2.S_1$, in addition to the previous two. This path also starts from a user node and ends in $S_1$, but it goes through $FW_2$ instead of $FW_1$. Note that $FW_2$ is also a firewall. So we refine our initial guess of the \intent to $dstIP:\IPa, start: User, waypoint:Firewall, end:S_1$, i.e. packets originating from the user nodes destined to $\IPa$ will traverse a firewall node and reach server $S_1$.
If we repeat this process for the paths destined to $\IPb$ (the red arrows), we get $dstIP:\IPb, start:User, waypoint:Firewall, end:S_2$.

Our two guesses for the green and red paths only differ in the last hop ($S_1$ vs. $S_2$). Both nodes are servers. 
So if we combine the information from all the paths, we can say that for packets destined to the prefix $\IPa/31$ originated at a user node, the packet will traverse a firewall and end up in a server node: $dstIP: \IPa/31, start: User, waypoint: Firewall, end: Server$.

\subsubsection{Example 2: ISP network}


Figure~\ref{fig:as} illustrates another example network that resembles part of a very simple ISP. The routers in the blue cloud show the network of an Autonomous System (AS) under our administration connected to two other ASes, namely $AS_1$ and $AS_2$.
The red arrows show the forwarding path for packets received from $AS_1$ that are destined to a specific IP prefix $P$ observed in a snapshot of the data plane taken in the morning: $AS_1.R_1.R_2.R_5.AS_2$.

Now let's assume that the link $R_1-R_2$ goes down and the control plane installs a new path for packets destined to $P$, namely $AS_1.R_1.R_3.R_4.R_5.AS_2$ in the next snapshot taken in the evening.
By looking at these two paths obtained from the two data plane snapshots across time, we can say what has remained invariant between the two paths is that the path destined to $P$ received from $AS_1$ will hit $R_1$, traverse some internal nodes and reach $AS_2$ through $R_5$. This can be denoted as $snapshot : Any, dstIP : P, path : AS_1.R_1.Internal^+.R_5.AS_2$ where $^+$ denotes $\geq$1 repetitions. 

\subsection{Applications}
\label{sec:applications}
Here we discuss some of the main applications of \intent inference. 

\textit{Enabling intent-based networking}:
The immediate application of the inferred \intents is to produce inputs for network verification and synthesis tools as, even if known, it is a tedious task to manually provide all \intents for a large legacy network. 
Generally, inferred intents can be used as input to any intent-based networking tool and streamline new applications including automatic migration from legacy networks to SDN or cloud paradigms, transparent network optimizations~\cite{oreo}, automatic network repair~\cite{neat}, etc.  

\textit{Network behavior summarization}: Current network management relies heavily on human in the control loop. Consequently, human insight is fundamental for network debugging~\cite{net2text} and management in general. Due to the complexity of large networks, an automatic tool to summarize relevant network behavior and present it in a comprehensible form can greatly help the process. 

\textit{Anomaly detection}: 
Another important application is to detect anomalies in forwarding behavior as a way to detect data plane or control plane bugs or misconfigurations. This is in contrast with traffic based anomaly detection techniques~\cite{anom1,anom2,anom3,anom4} for intrusion and DoS attack detection purposes.   


\name is useful in either application, though we focus on the first two in this paper (Sec.~\ref{sec:discuss}). 
Our framework is also more general and more expressive than the related works that solely focus on summarization~\cite{net2text} or network invariant inference~\cite{patent, config2spec} (Sec.~\ref{sec:related-work}). 

\section{\name Framework}
Using the classic precision/recall notions, we first define objective quality measures for intent inference (\ref{sec:overview}).
We then provide a framework to express network behavior and intents via features with hierarchical values to control the precision-recall trade-off, fitting the hierarchical nature of networks. We use it to formally define intent inference as an NP-hard constrained cost optimization problem related to our quality measures (\ref{sec:formal}).
We heuristically solve the problem by grouping relevant behavior using clustering techniques with the cost function as a distance measure and finding the most specific intent that represents all behavior per each group (\ref{sec:solution}).    


\subsection{Measures of quality}
\label{sec:overview}
\begin{figure}
    \centering
    \includegraphics[height=0.7in]{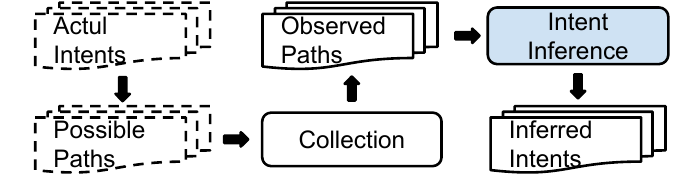}
    \vspace{-0.3cm}
    \caption{An abstract view of intent inference process}
    \label{fig:quality-overview}
    \vspace{-2em}
\end{figure}

In our abstract view of the intent inference process (Fig.~\ref{fig:quality-overview}), there is a set of \textit{actual intents} that govern the network behavior. Applied to the target network, the intents allow a set of \textit{possible} forwarding behavior in the form of a set of forwarding paths in the network.\footnote{Here, we assume a white-listing model meaning that any path not explicitly allowed by any intent in a set of intents is disallowed by that set.}
A \textit{collector} then collects a subset of these paths by data plane or control plane configuration analysis~\cite{deltanet, pec, plankton20, minesweeper}, observing the actual network traffic, etc. The collection mechanism is orthogonal to our work but we emphasize that the collector may not be able to observe all possible paths. For example, a traffic-based collector may miss behavior not exercised by the traffic, or some possible behavior may only be visible during link failures (as in Example 2), etc. 
The \textit{observed} paths are then fed into an \textit{intent inference} tool which consequently generates a set of \textit{inferred intents}.

In this context, we define the quality measures of intent inference. For a given set $P$ of paths and a set $I$ of intents, we define the number of true positives ($TP_P$), false negatives ($FN_P$), and false positives ($FP_P$) to be the number of paths in $P$ that are represented (i.e. allowed) by at least one \intent from $I$, paths in $P$ that are not represented by any \intents from $I$, and paths that are represented by the \intents in $I$ that are not in $P$, respectively. This way we are able to objectively measure the quality of intent inference through 
$\textit{Precision}_P  \triangleq TP_P / (TP_P + FP_P)$ and 
$\textit{Recall}_P \triangleq TP_P / (TP_P + FN_P)$ which respectively correspond to the \textit{specifity} (exclusion of incorrect behavior) and \textit{coverage} (inclusion of correct behavior) of its results. Clearly, both metrics must be considered to assess quality.   

Our goal in this paper is to be able to represent \textbf{all} (observed) behavior with as much precision as possible. So we view the intent inference problem as the problem of inferring a set of intents with the recall of 100\% (on observed behavior) that maximize precision.  
We note a special use case of intent inference, called summarization, where the set of possible paths is equal to the set of observed paths (perfect observation) and for human readability, the number of inferred intents is limited by a parameter ($k$). Without the limit $k$, one can simply list all observed paths as the inferred intents and achieve perfect precision and recall. We also consider and evaluate our tool in cases of imperfect observations as well, where $k$ is interpreted as a parameter to avoid over-fitting and under-fitting. 

\subsection{Formal setup}
\label{sec:formal}
In \name, forwarding behavior is expressed through features. 
Each \textit{feature} captures some aspect of forwarding behavior.
Examples include packet header information (e.g. source/destination IP address, port, protocol), device information such as start and end points, waypoints, ingress, egress, entire forwarding paths, the observation timestamp, device or topology state (e.g link status), etc. In Example 1 we used a tuple of destination IP, start, waypoint, and end features. In Example 2 we used a tuple of time of observation, IP destination prefix, and entire forwarding path features. 

Each feature can have a set of possible feature values or \textit{labels} associated with it. 
One of the main insights of this paper is to capture the hierarchies among feature labels that naturally fit networking environments.
For example, $\IPa$ and $\IPb$ are both included in $\IPa/31$ which itself is a subset of, say, $10.0.1.0/24$. Also in Example 1, both $FW_1$ and $FW_2$ are Firewalls. 
By supporting hierarchical values, \name allows for finer grained precision-recall trade-off control, resulting in higher quality intents (Section~\ref{sec:eval}). 

\newcommand{\Ddc}{D_\mathit{dc}}
\newcommand{\Disp}{D_\mathit{isp}}
\newcommand{\Dac}{D_\mathit{ac}}
\newcommand{\Dft}{D_\mathit{ft}}
\newcommand{\Tuple}{\mathit{Tuple}}
\newcommand{\Flat}{\mathit{Flat}}

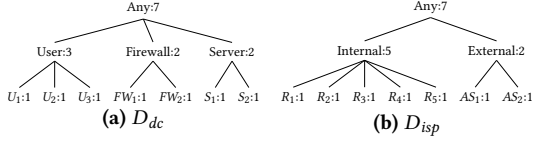
\begin{figure}[t!]
    \centering
    \captionsetup[subfigure]{aboveskip=-1pt,belowskip=-2em}
    \begin{subfigure}[t]{0.2 \textwidth}
        \centering
        \resizebox{\textwidth}{!}{
            \begin{tikzpicture}
            \Tree [.Any:7 [.User:3 $U_1$:1 $U_2$:1 $U_3$:1 ] [.Firewall:2 $FW_1$:1 $FW_2$:1 ] [.Server:2 $S_1$:1 $S_2$:1 ] ]
            \end{tikzpicture}
        }
        \caption{$\Ddc$}
        \label{fig:dc-label}
    \end{subfigure}%
    ~ 
    \begin{subfigure}[t]{0.2 \textwidth}
        \centering
        \resizebox{\textwidth}{!}{
            \begin{tikzpicture}
            \Tree [.Any:7 [.Internal:5 $R_1$:1 $R_2$:1 $R_3$:1 $R_4$:1 $R_5$:1 ] [.External:2 $AS_1$:1 $AS_2$:1 ] ]
            \end{tikzpicture}
            \vspace{-1em}
        }
        \vspace{-1em}
        \caption{$\Disp$}
        \label{fig:as-label}
    \end{subfigure}
    \caption{Device labeling hierarchy used in our examples}
    \vspace{-2em}
\end{figure}

To capture this, we formally define a \textit{feature type} $F$ as the tuple $((\Sigma_F,\subseteq), \delta_F)$ where $\Sigma_F$ is the set of possible labels for $F$ and $\delta_F: \Sigma_F \rightarrow  \mathbb{R}$ is a cost associated with each label.
We interpret the labels in $\Sigma$ as labeled sets of values which are partially ordered by the subset ($\subseteq$) relation. $F$ can essentially be represented by a DAG where nodes are labels in $\Sigma_F$ and edges are transitive reduction of the subset relation. 
Figures~\ref{fig:dc-label} and~\ref{fig:as-label} show example feature types for the set of devices used in Examples 1 ($\Ddc$) and 2 ($\Disp$), respectively. The number to the right of each label shows the cost of that label. For example $\delta_{\Ddc}(Firewall) = 2$.
Note how traversing from the bottom of these hierarchies to the top, the labels get less specific (lower precision) but cover more values (higher recall). 
We assign a higher cost to higher loss of precision (see below).

Any label of $\Sigma_F$ that is not a superset of any other label is called a \textit{concrete} label of $F$, i.e. concrete labels are the leaves of the DAG representing $F$ (denoted by $\sigma_F$). For label $l$,  $\sigma_F(l)$ denotes the subset of $\sigma_F$ included in the set labeled by $l$ -- e.g., $\sigma_{\Ddc}=\{U_1,U_2,U_3,FW_1,FW_2,S_1,S_2\} = \sigma_{\Ddc}(Any)$,$\sigma_{\Ddc}(Server)=\{S_1,S_2\}$.

A \textit{feature} is simply an instance of a feature type with a name.\footnote{When it is clear from the context, we use feature and feature type interchangeably.} 
We provide a library of features types that can be used (Section~\ref{sec:library}) to encode forwarding behavior. One can also design additional feature types according to the definition above. 
The library includes a feature type for a tuple of multiple other feature types. We used the feature $(dstIP,start,waypoint,end)$ as an instance of $\Tuple \langle  \mathit{IPPrefix}, \Ddc, \Ddc, \Ddc \rangle$ in Example 1. 

Within this setup, for a feature $F$, a \textit{path} is simply a value from $\sigma_F$ and an \textit{intent} is a value from $\Sigma_F$. 
For a set of intents $I = \{i_1,...,i_k\}$ and a set of paths $P = \{p_1,...,p_n\}$ we say $I$ represents $P$ iff  $P \subseteq \bigcup_{i \in I} \sigma_F(i)$. For instance $(\IPa,User,Firewall,S_1)$ represents $\{(\IPa,U_1,FW_1,S_1), (\IPa,U_3,FW_2,S_1)\}$.

We formalize the intent inference problem in this framework:
\begin{definition}[Intent inference problem]
\label{def:intent-inference-problem}
For a given feature $F$, a set of paths $P$, and a limit on the number of inferred intents ($k$), find the set of intents $I^* = \{i_1,...,i_{k'}\}$ ($k' \leq k$)  (the \textit{inferred \intents}) that represents $P$ and minimizes $\delta_F(I^*) = \sum_{i \in I^*} \delta_F(i)$. 
\end{definition}

For example for feature $\Ddc$, $P = \{U_1,U_3,S_1\}$, for limits of 3, 2, and 1, it is easy to see the set of inferred intents are $\{U_1,U_3,S_1\}$, $\{User,S_1\}$, and $\{Any\}$ with the costs of 3, 4, and 7, respectively. 

To understand the relation between this definition and the intuition provided in the last section, note that if we set the cost of each label to the number of concrete values it represents (as we mostly do in our feature library\footnote{See Sec.~\ref{sec:library}. One can also alter the costs to guide the inference.}), for any $I$ in the set of all sets of intents representing $P$ ($\mathbb{I}_P$), $\delta_F(I)$ approximates the number of paths that the intents in $I$ collectively represent, i.e. $TP+FP$\footnote{The imprecision is due to over-counting overlapping intents. We penalize overlap to encourage inferring disjoint intents.}. Also, note that $TP$ is the same for any such $I$. So $\delta_F(I)$ is inversely related to the precision of $I$. Also for any such $I$, recall (on observed paths) is 1. 

It is possible to show the intent inference problem as defined above is NP-hard by a reduction from the set cover problem. 
We provide a heuristic polynomial solution to the problem in Section~\ref{sec:solution}. 

\vspace{-1em}
\subsection{Feature types library}
\label{sec:library}
Inspired by~\cite{pec}, we provide a template library of feature types.
Depending on the type of network, collection mechanism, intended application, domain-knowledge, etc, different type templates in the library can be instantiated to encode the forwarding behavior. One can also design additional feature types.
The following are examples of templates supported in our library: 

$DAG \langle V,E \rangle$ where $V$ and $E$ are nodes and edges of an arbitrary DAG, is a feature type where any label $v \in V$ is interpreted as the set of leaves reachable from $v$. $\Ddc$ is an example of this type.  

$\Flat\langle S \rangle$, where $S$ is a set of concrete values, is a feature type defined by a DAG with a single root ($Any$) connected to $|S|$ leaves each corresponding to a member of $S$. An intent can either be a concrete value or \textit{anything}, with no hierarchy in between. In the ISP example, we used $\Flat\langle \{Morning, Evening\} \rangle$. 

$TBV\langle n \rangle$ is a feature defined over Ternary Bit Vectors (TBV) of length $n$. A TBV is a generalization of bit-vectors where arbitrary bits can be wildcards. A TBV is interpreted as the set of bit-vectors it represents, e.g. $0x \equiv \{00,01\}$. $\mathit{IPPrefix}$ is a specialization of $TBV\langle 32 \rangle$ where wildcards can only happen at the end of TBVs.

$Range$. Integer ranges form a feature type suitable for fields like IP and port ranges or constraints like number of link failures.  


$HRE\langle F,d \rangle$ is a variant of regular expressions over hierarchical alphabet useful for representing entire forwarding paths (Sec.~\ref{sec:hre}).

$\Tuple \langle F_1, ..., F_n \rangle$ is used to combine multiple features to create more complex features. $\Sigma = \Sigma_{F_1} \times ... \times \Sigma_{F_n}$, and for any $a,b \in \Sigma$, $a \subseteq b$ iff $\bigwedge_{i \in [1,n]} a_i \subseteq b_i$. Finally $\delta(a) = \prod_{i \in [1,n]} \delta_{F_i}(a_i)$. 

For all of these features (except \hre) $\forall l \in \Sigma : \delta(l) = |\sigma(l)|$. 

\vspace{-0.7em}
\subsubsection{Representing entire forwarding path}
\label{sec:hre}
As an exemplar of a complex feature, we describe a feature type designed to represent entire forwarding paths, such as the ones used in the ISP example. 

Our idea is to use regular expressions to represents sets of such paths. 
Specifically, we focus on a very limited class of regular expressions that seems to be a proper fit for representing network paths.
The grammar of such regular expressions is shown in Figure~\ref{fig:grammar}. An \hre is defined over another feature type $F$ and its alphabet is $\Sigma_F$. We slightly generalize the notion of acceptance in our regular expressions to account for the hierarchy of labels we introduced. We call these \textit{Hierarchical Reduced Regular Expressions} (\hre). 

Here, a path is a string over $\sigma_F$. We say a path $p$ is represented by an \hre $h$ iff there exists a string $s$ obtained by replacing a subset of the labels in $p$ with another label from the set of ancestors of that label in $\Sigma_F$ and $s$ is accepted by $h$ interpreted as a normal regex. We denote by $Acc^d_F(h)$ the set of all strings over $\sigma_F$ of length $\leq d$ represented by $h$. 
In the ISP example, the path $p = AS_1 . R_1 . R_3 . R_4 . R_5 . AS_2$ is accepted by the \hre $h = AS_1 . R_1 . Internal^+ . R_5 . AS_2$ because $s = AS_1 . R_1 . Internal . Internal . R_5 . AS_2$ which is obtained by replacing $R_3$ and $R_4$ in $p$ by the label $Internal$ (ancestor to both device names in $\Disp$), is accepted by interpreting $h$ as a normal regex.
For a feature type $F$, and a limit $d$ on length of strings, we define $HRE\langle F,d\rangle$ as $((\Sigma_H,\subseteq), \delta_H)$ where $\Sigma_H$ is the set of all \hres over $\Sigma_F$ with length $\leq d$. We interpret each $h \in \Sigma_H$ as the set $Acc^d_F(h)$ and the hierarchy is formed according to the subset relation among these sets, e.g. $h_0 = AS_1 .  R_1 . R_2 . R_5 . AS_2 \subset h_1 = AS_1 . R_1 . Internal . R_5 . AS_2 \subset h_2 = External . Internal^+ . External \subset h_3 = Any^+$ ($d \geq 5)$. 

By the argument in Sec.~\ref{sec:formal} we should set $\delta_H(h) = |\sigma_H(h)| = |Acc^d_F(h)|$. However, computing $|Acc^d_F(h)|$ is expensive. 
Instead, we roughly approximate the value by $m_F(h)^d$ where $m_F(h)$ is the the geometric mean of cost of labels of \hre $h$ over field $F$ (i.e. the average number of concrete labels of $F$ each label in $h$ represents).
In Example 2, $\delta_H$ for $h_0,...,h_3$ are $1^d$, $1.38^d$, $2.71^d$, and $7^d$, respectively. Note how less precise intents received higher costs. 


\begin{figure}[t]
    $\texttt{HRE} ::= l\ \ |\ \ l^+\ \ |\ \ \texttt{HRE}.\texttt{HRE}$  \hspace{1cm} $l \in \Sigma_F
    $
    \vspace{-0.4cm}
    \caption{\small{Grammar of Hierarchical Reduced Regular Expressions}}
    \label{fig:grammar}
    \vspace{-2em}
\end{figure}

\vspace{-0.7em}
\subsection{Solving the intent inference problem}
\label{sec:solution}
Our heuristic solution to the intent inference problem divides it into two related sub-tasks: single intent inference and path selection. 
\vspace{-0.7em}
\subsubsection{Single intent inference}
This is a specialization of the intent inference problem with $k = 1$: find a single intent that represents all input paths with the lowest cost.
For a feature $F$, we define the function $\sqcup_F: 2^{\sigma_F} \rightarrow \Sigma_F$ (called the \textit{join} function) as the answer to the single intent inference problem. In practice (Sec.~\ref{sec:cluster}) we use $\sqcup_F: \Sigma_F^2 \rightarrow \Sigma_F$ where $\sqcup_F(a,b) = \sqcup_F(\sigma_F(a) \cup \sigma_F(b))$.

For most of the feature types in our library, computing join of two labels is straightforward and efficient: $\sqcup_{DAG}(a,b)$ is their least cost common ancestor, $\sqcup_{\Flat}(a,b)$ is $a$ if $a = b$ else $Any$, $\sqcup_{TBV}(a,b) = c$ where $c_i = a_i$ if $a_i = b_i$ else $c_i = x$. $\sqcup_{Range}([a,a'],[b,b']) = [min(a,b),max(a',b')]$, $\sqcup_{\Tuple}(a,b) = (\sqcup_{F_1}(a_1,b_1),...,\sqcup_{F_n}(a_n,b_n))$.
Computing join for $HRE\langle F, d \rangle$ is more complex and requires dynamic programming ($O(d^3|\Sigma_F|^2)$). We omit the details due to space limits.  
For example, $\sqcup_{\Ddc}(U_1,U_2)=User$, $\sqcup_{\Ddc}(U_1,FW_2)=Any$ and the join of the two paths in Example 2 is $AS_1. R_1. Internal^+. R_5 . AS_2$

\textit{Observation}: Because of the way we defined label costs, we can use $\delta(\sqcup(P))$ as a measure of similarity/relatedness of the paths in a set $P$. Higher cost means we have to lose more precision to represent all paths together using a single intent, hence the paths are probably not related to each other, i.e. do not come from the same intent.  


\vspace{-0.7em}
\subsubsection{Path selection}
\label{sec:cluster}
Having a solution for single intent inference, the next task is to decide which subset of paths should be fed into the single intent inference problem. 
Following the observation above, we treat the general intent inference problem roughly as a clustering problem where the goal is to put the more similar paths into the same clusters and then feed these clusters as inputs to the single intent inference problem to infer one intent per cluster.
In our clustering, the similarity measure mentioned above can be used to define the distance between paths and clusters of paths. 

Specifically our clustering approach is inspired by the Hierarchical Agglomerative Clustering (HAC) technique~\cite{hac}. In HAC, each object (i.e. path) is considered as a separate cluster initially. The clusters are then iteratively merged with each other until a single cluster remains. At each iteration, the clusters with the minimum distance are selected to be merged with each other. In our approach we terminate the iteration once at most $k$ clusters remain. We define the distance $d(c_i, c_j)$ between clusters $c_i$ and $c_j$ as the amount of increase in the cost of representation by merging the clusters compared with the sum of cost of individual clusters: $d(c_i, c_j) \triangleq \delta(\sqcup(c_i \cup c_j)) - (\delta(\sqcup(c_i)) + \delta(\sqcup(c_j)))$.


As an optimization we approximate $\sqcup(c_i \cup c_j)$ by $\sqcup(\sqcup(c_i), \sqcup(c_j))$. $\sqcup(c_i)$ and $\sqcup(c_j)$ are approximated recursively during clustering. This way, the join function is applied to only two labels at a time. 
As another optimization, after forming each new cluster $c_n$, instead of computing the distance of $c_n$ to all other clusters, we only compute it for a random subset of them with a configurable size $b$. 
With these optimizations, the complexity of our solution for $N$ paths is $O(JbN\mathit{log}N)$ where $J$ is the complexity of join (for two inputs).

\begin{figure*}[t!]
    \centering
    \captionsetup[subfigure]{aboveskip=3pt,belowskip=-2em}
    \begin{subfigure}[t]{0.20 \textwidth}
        \includegraphics[width=\textwidth]{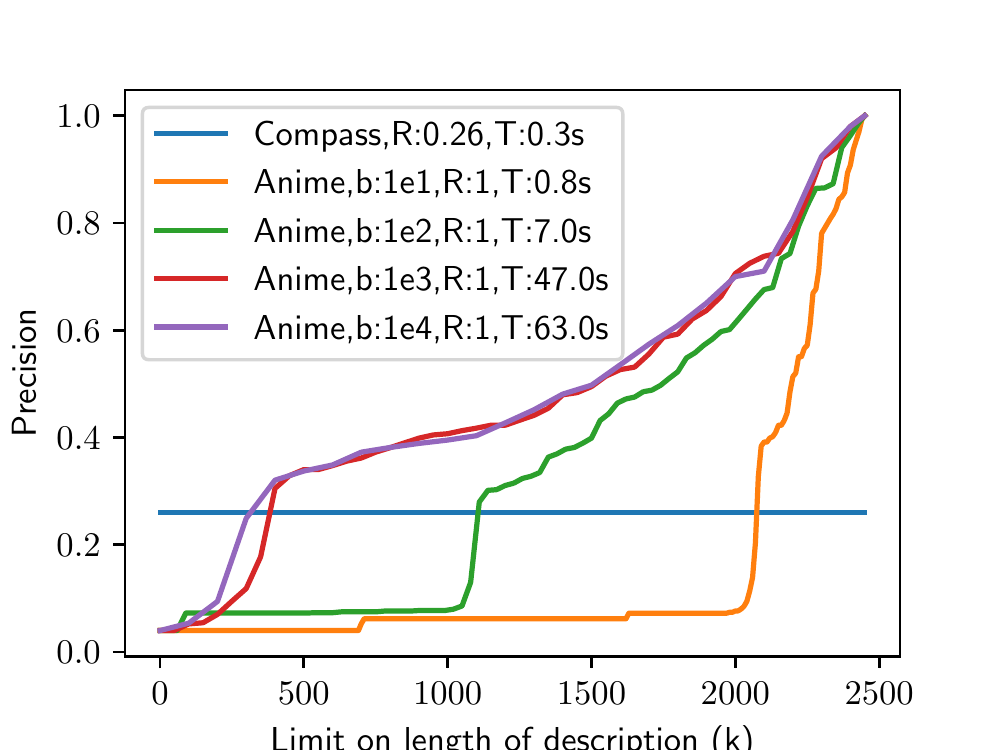}
        \caption{ATT, 100\% observation}
        \label{fig:att100-precision}
    \end{subfigure}%
    ~ 
    \begin{subfigure}[t]{0.20 \textwidth}
        \includegraphics[width=\textwidth]{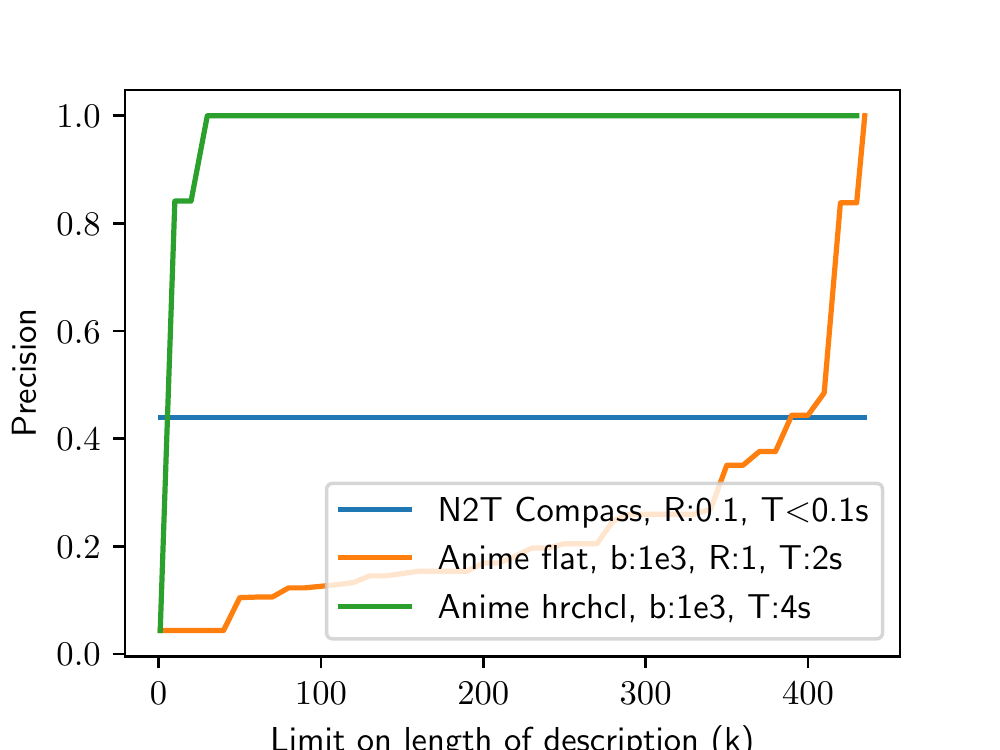}
        \caption{\small{Access cntrl., 100\% obsrv.}}
        \label{fig:ac-precision}
    \end{subfigure}
    ~
    \begin{subfigure}[t]{0.20 \textwidth}
        \includegraphics[width=\textwidth]{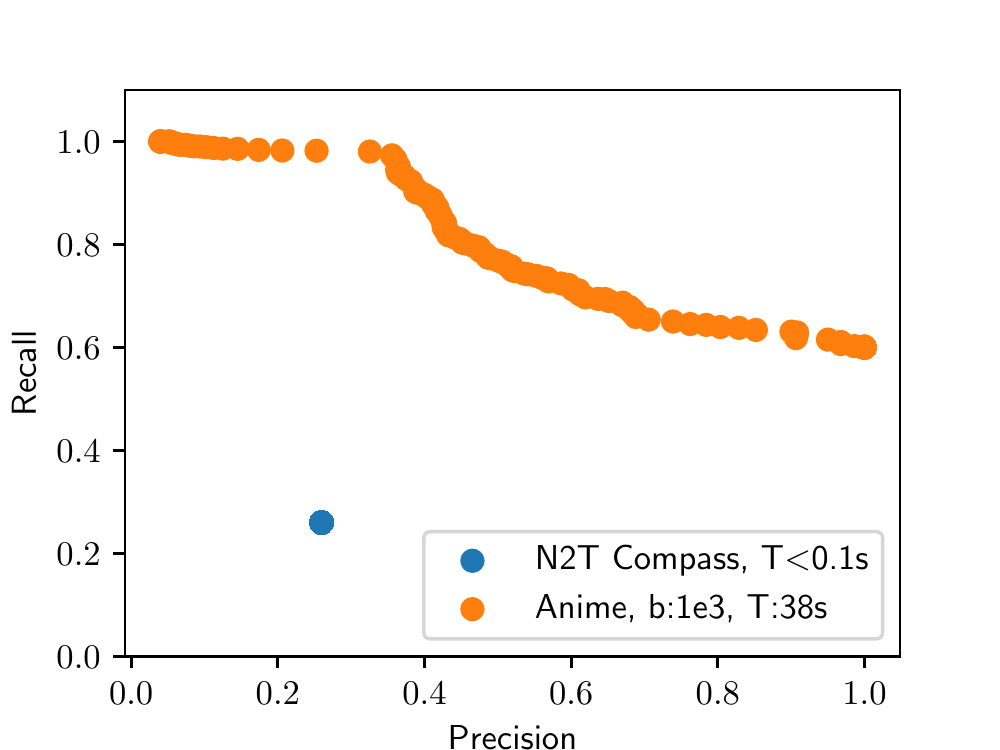}
        \caption{ATT, 60\% observation}
        \label{fig:att100-60-p-r}
    \end{subfigure}
    ~ 
    \begin{subfigure}[t]{0.20 \textwidth}
        \includegraphics[width=\textwidth]{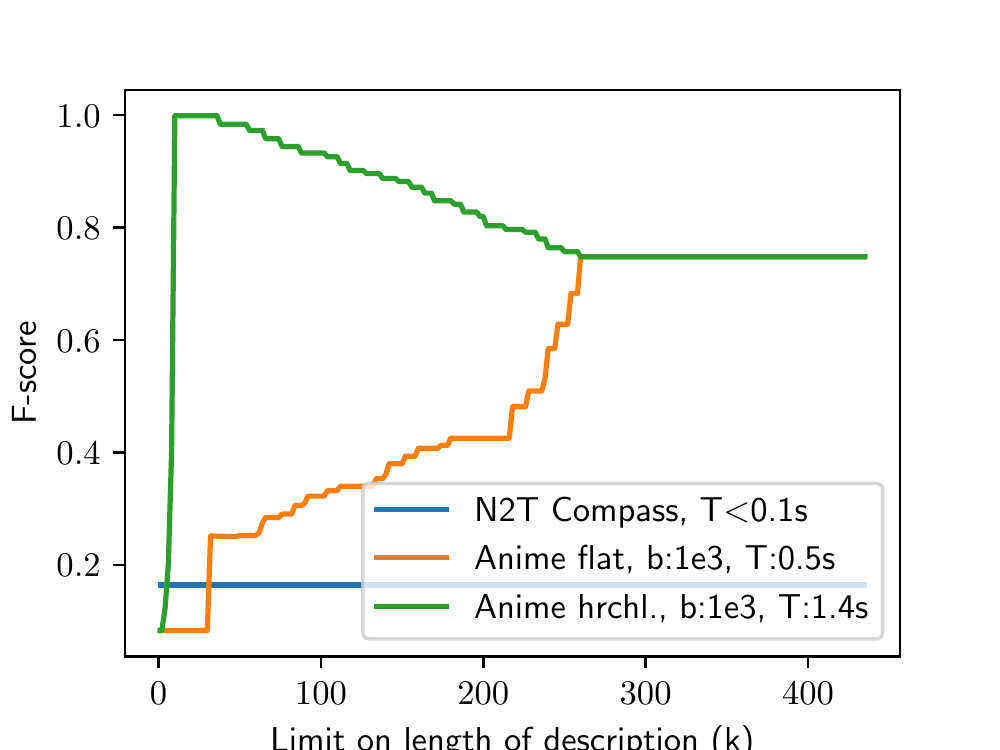}
        \caption{\small{Access cntrl., 60\% obsrv.}}
        \label{fig:ac-60-fscore}
    \end{subfigure}%
    ~
    \begin{subfigure}[t]{0.20 \textwidth}
        \includegraphics[width=\textwidth]{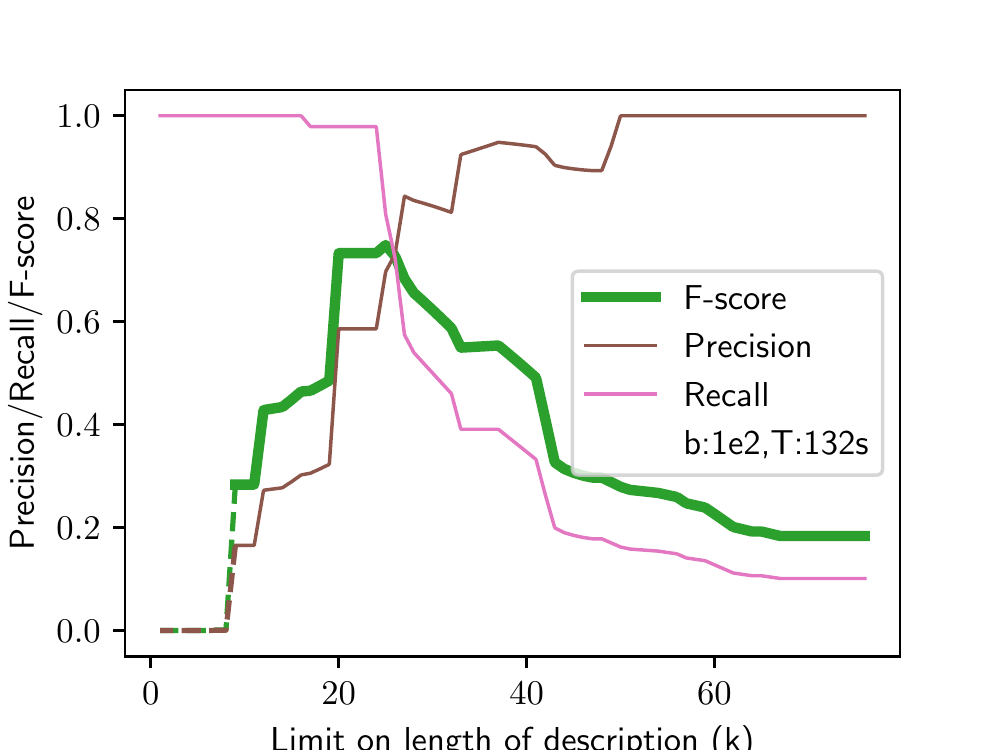}
        \caption{\small{DC HRE., 10\% obsrv.}}
        \label{fig:hre}
    \end{subfigure}%

    \caption{\name experimental results. In the legends, $b$, $R$, and $T$, stand for the batch size, recall rate, and average runtime.}
    \vspace{-1em}
\end{figure*}

\vspace{-1em}
\section{Evaluation}
\label{sec:eval}
We implemented a prototype of our framework in ~1K lines of Python code. As our baseline, we re-implemented Compass, a heuristic algorithm for Net2Text's~\cite{net2text} formulation of the network summarization problem (Sec.~\ref{sec:related-work}).
In our experiments, we assess the performance of \name and the quality of its inferred \intents and compare it with Compass. Particularly, we study the effect of the clustering algorithm, parameters $b$ and $k$, flat vs. hierarchical labeling, \hres, and perfect vs. imperfect observation. All experiments ran on a Macbook Air, 1.6GHz Intel Core i5, 8GB 1600MHz DDR3 RAM.
\vspace{-0.7em}
\subsubsection{Comparison with Net2Text}
We reuse Net2Text's own evaluation dataset used in~\cite{net2text}. The dataset uses real-world topologies, IPv4 RIB, and AS-to-organization information. It simulates the forwarding state in a simplified ISP network. Each dataset entry contains various information about a path traversing through the ISP from an ingress to an egress. 
Because Net2Text can not effectively deal with entire forwarding paths or hierarchical values (Section~\ref{sec:related-work}), we only focus on the ingress device and egress devices and the destination organization of each path to have a fair comparison. Specifically, we use the feature $(organization,ingress,egress)$ as an instance of $\Tuple\langle \Flat \langle O \rangle, \Flat \langle D \rangle, \Flat \langle D \rangle \rangle \rangle$ where $D$ is the set of network devices and $O$ is the set of organizations.
We experimented with various topologies with varying nodes, egresses, and destinations. Figure~\ref{fig:att100-precision} shows a representative result using the AT\&T topology with 25 nodes, 5 egresses, and 100 destinations, resulting in near 2500 paths. 
The x and y axes show the limit on the number of inferred intents ($k$) and the precision, respectively. Recall rate (same for all $k$) and average runtime (over all $k$) are reported in the legend.
As expected, \name's precision increase as we increase $k$. The batch size ($b$) has a great impact on the rate of increase in precision (and performance). In any case, \name's recall is 100\%. This is due to the fundamental design decision of representing all observed behavior, which is encoded in our problem definition and approach. 
Net2Text's Compass algorithm has a fixed precision and recall of near 26\%. By increasing $k$, Compass produces new intents that are subsets of previous intents, thus no increase in precision or recall. Compass tries to optimize Net2Text's scoring function which is designed according to Net2Text's goals and assumptions (Section~\ref{sec:related-work}), not directly the precision and recall, although the scoring function is related to these two factors in an ad-hoc way.

\vspace{-0.7em}
\subsubsection{Effect of hierarchies}
To show the effects of hierarchical values on the quality of inferred intents, we use a synthetic dataset resembling simplified network access control policies. For a network with $n$ endpoints (assuming each is a servers in a data-center), we randomly partition the servers into $g$ groups of random size between $min$ and $max$. All groups are subsets of $Any$.  
We then randomly generate $m$ intents each of the form \textit{``server or group $a$ can communicate with server or group $b$''}. We generate the set of all server pairs represented by the intents as input for our experiment. 

We experiment with two types of features for \name, namely $\Tuple\langle \Flat \langle D \rangle, \Flat \langle D \rangle \rangle$ and $\Tuple\langle \Dac, \Dac \rangle$ where $D$ is the set of servers and $\Dac$ is the feature type defined by the DAG representing the hierarchy we described above. We also test Compass. 

Figure~\ref{fig:ac-precision} shows the result of a representative experiment with $n = 100, g = 5, min = 5, max = 30, i = 10$ resulting in $435$ paths. \name with flat label has worse precision than Compass but 10x better recall. However, hierarchical labeling guides \name to achieve both good  precision and recall (even for small values of $k$).
\vspace{-0.7em}
\subsubsection{Imperfect observations}
We assess \name's effectiveness in the face of imperfect observations by repeating the previous experiments with random subsets of the original input. We then compute precision and recall according to the original input. In this setting, $k$ is interpreted as parameter one can tune to avoid over/under-fitting.

Figure~\ref{fig:att100-60-p-r} shows the precision (x-axis) recall (y-axis) tradeoff for different values of $k$ in the ATT experiment where only 60\% of the original paths are observed. As expected, an increase in $k$ results in higher precision and lower recall rates. \name strikes a better balance between precision and recall than Compass. 

Figure~\ref{fig:ac-60-fscore} shows the results of similar experiments with the access control dataset. The x-axis shows $k$ and the y-axis shows the F-score of the result which is defined as the harmonic average of precision and recall. 
As $k$ gets smaller, hierarchical labeling achieves better recall without scarifying precision thus achieves better F-score. Higher recall for flat labeling significantly sacrifices precision, thus its F-score declines.  Hierarchical labeling achieves near-perfect F-score (1 FN, 0 FP) for $k \in [10,26]$ where 9/10 of the actual intents are inferred. For $k < 10$, hierarchical labeling starts to underfit and its precision and F-score fall sharply. For the optimal $k$, with either labeling, \name achieves >4x better F-score than Compass. 

\vspace{-0.7em}
\subsubsection{Experiment with \hres}
To showcase the use of \hres we create a synthetic data center topology consisting of $c$ clusters, each containing $f$ firewalls connected to $p$ spine switches which are themselves connected to $l$ leaf switches. Each leaf connects $r$ racks, each containing $s$ servers. The firewalls are connected to $g$ gateway routers shared among the clusters. The gateways are connected to $i$ ISPs providing Internet connectivity\footnote{The full topology and hierarchy of device labels are available in~\cite{dc-hre-results}.}. We consider the following actual intents: The servers within each cluster can talk to each other and to the internet. Internet can talk to the servers in a special cluster called DMZ. DMZ cluster servers can talk to servers in any other cluster. The set of possible paths are all shortest paths allowed by the intents. We take a subset of possible paths where only one random path among all allowed paths between any two points is observed. We feed the observed paths to \name with feature $HRE\langle \Dft, 8 \rangle$ where $\Dft$ is the hierarchy described above.
We run an experiment with $c,f,p,s,g,i=2;r=1$ resulting in ~75 observed paths (750 possible paths). Below is an example output for $k = 9$ achieving precision and recall of 20\% and 100\%, respectively. 
\normalsize

\begin{tiny}
\begin{verbatim}
1:Server.Leaf.Spine.Firewall.Gateway.Internet, 
2:Internet.Gateway.DMZFirewall.DMZSpine.DMZLeaf.DMZServer,
3:DMZServer.DMZLeaf.DMZSpine.DMZFirewall.Gateway.Cl1Firewall.Cl1Spine.Cl1Leaf.Cl1Server,
4:Cl1Server.Cl1Leaf.Cl1Spine.Cl1Leaf.Cl1Server, 5:Cl1Server.Cl1Leaf.Cl1Server, 6:Cl1Server,
7:DMZServer.DMZLeaf.DMZSpine.DMZLeaf.DMZServer, 8:DMZServer.DMZLeaf.DMZServer, 9:DMZServer
\end{verbatim}
\end{tiny}
\normalsize
\vspace{-0.5em}
The imprecision is due to intents partly representing non-optimal or impossible paths, e.g. 4 includes paths between servers connected to the same leaf that go through a spine. 
Figure~\ref{fig:hre} shows the precision, recall, and F-score of the results. 
F-score is near its peak value (0.75) for $k \in [20,25]$, and falls as we move away from that range. 

\vspace{-0.7em}
\subsubsection{Performance}
\name's performance is significantly influenced by (1) value of $b$, (2) number of input paths, and (3) complexity of its features. (1): In the ATT experiments, 10x and 100x larger values of $b$ (compared to $b=10$, $0.3s$) incur 20x ($7s$) and 150x ($47s$) slowdown, respectively. 
(2): For the same $b$, $40\%$ reduction in input size corresponds to $21\%$ speedup ($38$ vs $47s$) in ATT. 
(3): 
The hierarchical feature caused 2x ($4s$ vs $2s$) slowdown in access control experiment. Also the data center \hre experiment is 1.2K times ($2m$ vs. $0.1s$) slower than an experiment with the same input size and $b$ for ATT as \hre join is computationally very expensive. 
Finally, \name's performance is only slightly influenced by (4) the value of $k$ -- e.g. $55s$ vs. $34s$ for min and max $k$ in ATT, $b=1000$.

Compass is faster than \name as it is almost linear in the number of inputs while \name is almost quadratic for large values of $b$ -- e.g. 0.3s vs 46s in ATT, $b$=$1000$. Still, \name's performance is acceptable given its intent inference quality. Here we focus on the intent inference framework itself and leave performance optimizations (e.g. by using other clustering techniques) as future work. 

\vspace{-1.4em}
\section{Discussion}
\label{sec:discuss}

\textit{Expressiveness.} 
Our formalism can express/infer important classes of functional intents incl. reachability and waypointing under various (temporal, topological, header, etc.) conditions. 
It is not designed to explicitly infer negative behavior such as in isolation intents. This can be alleviated by encoding negative behavior as positive behavior, e.g. packet drop as reachability to a special node, or using labels representing complemented sets (e.g non-firewall or $!\{\texttt{TCP},\texttt{UDP}\}$). 

\textit{Feature engineering and parameter tuning}
We do not necessarily expect the end-user (network operator) to be in charge of these tasks. 
A front-end layer between Anime and the end-user -- designed for specific application, network type, collector, etc. -- can abstract the low-level details, though the user may be given direct/indirect (see below) control for better results. The layer can also employ automatic techniques for feature selection/parameter tuning -- e.g. in summarization, it can suggest promising values for $k$ using the elbow method; for prediction, $k$ can be tuned via cross-validation.  

\textit{Incorporating user feedback.} 
An interesting future direction is to make the inference process more interactive: the system proposes some intents, and the user selects intents that make sense or marks the ones that do not seem correct. She may also directly provide a \textit{negative intent}, which can be used as explicit negative labels for the paths represented by it. The system then infers new intents that respect the user feedback for example by tuning label costs, etc. 

\textit{Faulty behavior.}
\name can be used to detect faulty behavior by summarizing all forwarding behavior in a human-comprehensible form. User can inspect the results for anomalous behavior. While we did not address \textit{automatic} anomaly detection, our framework is useful for designing such a method, a major future work direction.

\textit{Real-world experiments and user study.} We leave larger scale and more real-world experiments as future work. In addition, although we have devised objective measures of quality for intent inference, the perceived quality is determined by the end-user. Therefore, another important direction for future work is to perform user studies on the usefulness of \name for network operators.

\vspace{-1em}
\section{Related Work}
\label{sec:related-work}
Nex2Text addresses the problem of summarizing network traffic ``as much as possible''~\cite{net2text}. 
Net2Text's formal setup can be thought of as a special case of \name's where the only allowed feature type is of the form $\Tuple \langle \Flat \langle S_1 \rangle, ..., \Flat \langle S_n \rangle \rangle$.
The authors assign a \textit{score} to each summary (= set of inferred intents in \name) that essentially awards representing more paths/traffic and using labels other than $Any$. The score is proportional to precision and recall in an ad-hoc way. 
The paper provides an approximate algorithm (Compass) that, for a limit on the length of description, produces a summary with max score.
Unlike \name, Net2Text cannot handle hierarchies. 
As demonstrated in Sec.~\ref{sec:eval}, this limitation significantly affects the quality of inferred intents. 
Net2Text cannot handle entire paths either. These limitations prevent Net2Text from effectively representing \intents such as the ones in our motivating examples.  


\name is also related to a work~\cite{patent} aimed at finding forwarding invariants 
by observing the reachability relation among network devices for each header equivalence class~\cite{deltanet} and intersecting this relation over snapshots of the network obtained over time (e.g. after link failures). 
The same basic idea has also recently been used in Config2Spec~\cite{config2spec}. These works are inspired by software spec. mining literature, particularly dynamic invariant inference~\cite{daikon}.
Our work is a generalization of this idea in the sense that in addition to considering invariants over time/snapshots, we also consider invariants over other dimensions such as network devices, packet header fields, etc. and infer the ones that best express all observations across all dimensions. E.g.~\cite{patent,config2spec} cannot infer any higher-level information from a single snapshot like in Example 1, while \name can.


\vspace{-1em}
\section{Conclusion}
\name framework enables a novel approach towards bridging the semantic gap between high-level network intents and low-level behavior by inferring the former from the later. 
Our experiments on various datasets demonstrate the effectiveness of our approach in inferring high-quality intents with acceptable performance.

\textbf{Acknowledgements.}
We especially thank Prof. Brighten Godfrey for his great guidance. We also thank Prof. Madhu. Parthasarathy, Prof. Matthew Caesar, Santhosh Prabhu, our shepherd Muhammad Shahbaz, and the anonymous reviewers of SOSR for their comments and suggestions. 
This work is supported by NSF grant CNS-1513906.

\bibliographystyle{ACM-Reference-Format}
\bibliography{main}

\end{document}